# A further study on the high-precision optics for HIAF-BRing

Ke Wang [a, b, c, d], Li-Na Sheng [b, c], Tao Li [b, c, e], Geng Wang [b, c], Wei-Ping Chai [b, c], You-Jin Yuan [b, c], Jian-Cheng Yang [b, c], Guo-Dong Shen [b, c*], Liang Lu [a, d†]

a. Sino-French Institute of Nuclear Engineering and Technology, Sun Yat-sen University, Zhuhai 519082, P. R. China

b. Institute of Modern Physics, Chinese Academy of Sciences, Lanzhou 730000, P. R. China

c. University of Chinese Academy of Sciences, Beijing 100049, P. R. China

d. United Laboratory of Frontier Radiotherapy Technology of Sun Yat-sen University & Chinese Academy of Sciences Ion Medical Technology Co., Ltd, Guangzhou, P. R. China

e. School of Nuclear Science and Technology, Lanzhou University, Lanzhou 730000, P. R. China

**Abstract**

The High Intensity heavy ion Accelerator Facility (HIAF) successfully accelerated the $^{18}O^{6+}$ beam on October 27, 2025. This paper presents a further simulation study on the high-precision optics, namely sliced optics, of the Booster Ring (BRing) at HIAF based on measured magnetic fields, focusing on three aspects: (1) closed-orbit distortion (COD) and variations in optical parameters induced by errors; (2) closed-orbit correction; (3) dynamic aperture. Specifically, detailed investigations are conducted on COD and optical parameter variations caused by magnet alignment errors and dipole magnet field errors, alongside simulations of closed-orbit correction and detailed calculations of BRing's dynamic aperture. Results show the sliced optics outperforms the original optics in COD control. Without chromaticity correction, its dynamic aperture is superior to the original; after chromaticity correction, it remains comparable. This study provides valuable insights for accelerator tuning and optimization.

## 1. Introduction

The High Intensity heavy ion Accelerator Facility (HIAF) [1-3] constructed by the Institute of Modern Physics, Chinese Academy of Sciences, has been completed. The Booster Ring (BRing) [4, 5] is the core of the entire facility. It is capable of accelerating $^{238}U^{35+}$ ions to 0.8 GeV/u with a beam intensity of 1.0E+11 ppp (particles per pulse). On October 27, 2025, BRing successfully accelerated $^{18}O^{6+}$ from 30

[*] Corresponding author: Guo-Dong Shen, Email: shenguodong@impcas.ac.cn.

[†] Corresponding author: Liang Lu, Email: luliang3@mail.sysu.edu.cn.

MeV/u to 2600 MeV/u. The next beam commissioning will prioritize optimizing the beam intensity. The lattice structure of BRing is FODO, consisting of 3 super-periods, each of which includes an arc section and a straight section, with a circumference of 569 m. Each arc section contains 16 dipole magnets and 16 quadrupole magnets, while each straight section houses 10 quadrupole magnets. A total of 79 correction magnets are installed for closed-orbit correction, including 39 horizontal correction magnets, 38 vertical correction magnets, and 2 bidirectional correction magnets. Forty-eight sextupole magnets are employed for chromaticity correction, divided into 2 groups with 24 sextupole magnets each. The elements layout is shown in Figure 1. The other important components of HIAF are mainly the superconducting electron cyclotron resonance ion source (SECR)[6], the superconducting linear accelerator (iLinac)[7, 8], the High energy FRagment Separator (HFRS) [9, 10], and the Spectrometer Ring (SRing) [11, 12]. The layout of HIAF is shown in Figure 2. Both the beam energy and intensity of HIAF are expected to reach the world's advanced level, which places high demands on accelerator tuning.

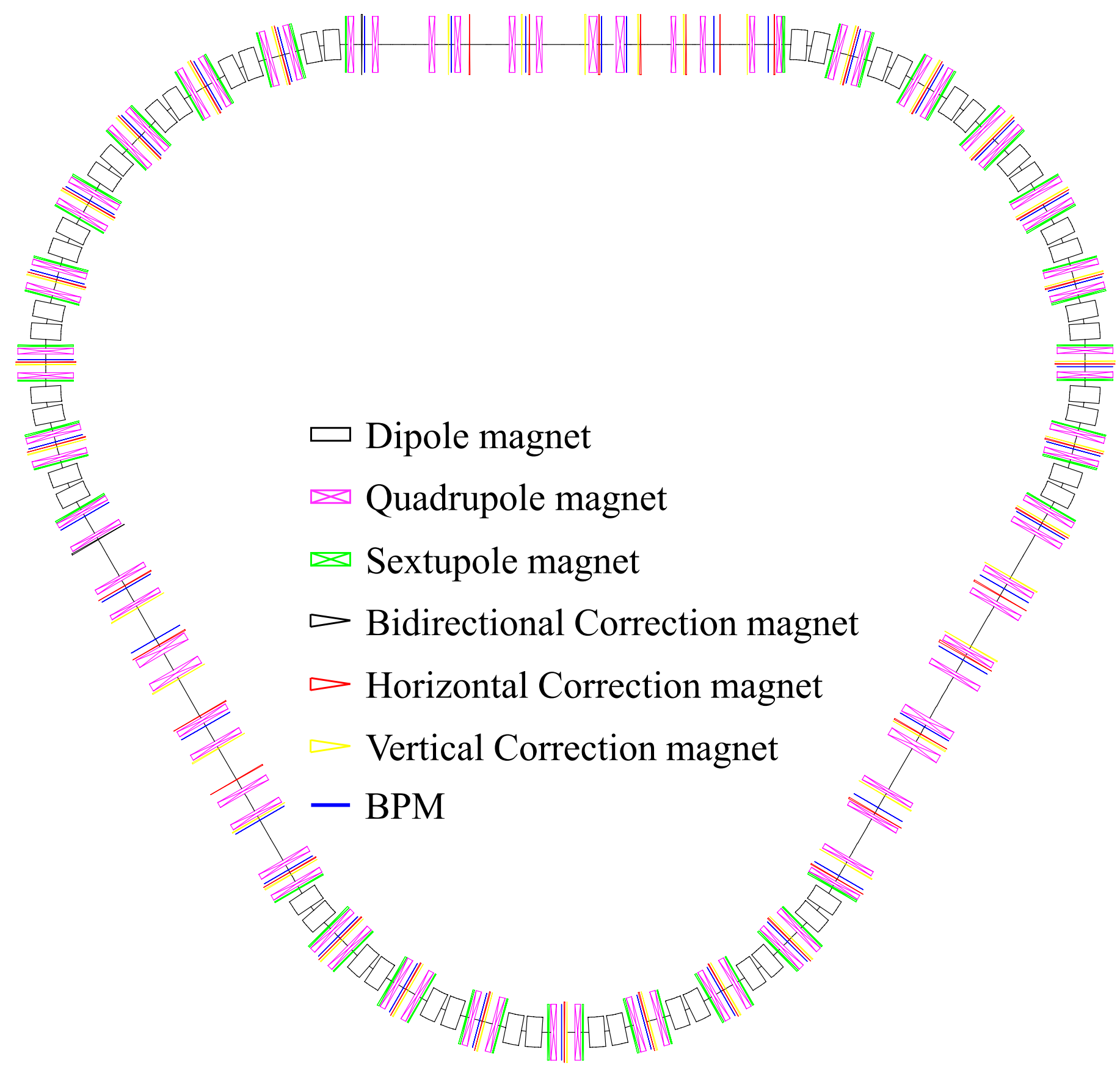

Figure 1. Schematic of BRing.

In the previous publication [5], a high-precision optical model of BRing, namely the sliced optical model or sliced optics, is established using the measured magnetic field data of magnets. In this sliced optical model, each magnet is divided into hundreds of segments based on the magnetic measurement data. This model not only accounts for the fringe field effect of magnets but also considers the magnetic field overlap between adjacent magnets, so it is more consistent with the actual operational conditions of the accelerator. Calculations show that under the premise of keeping the magnet longitudinal field integral unchanged, both the horizontal and vertical working points of the sliced optics are significantly reduced compared with those of the original optics. Additionally, the straight sections that are originally achromatic no longer maintain this achromatic property, so it is necessary to adjust the strengths of the quadrupole magnets. By rematching the strengths of the quadrupole magnets, the optical parameters of the sliced optics can be made consistent with those of the original optics, as shown in Figure 3.

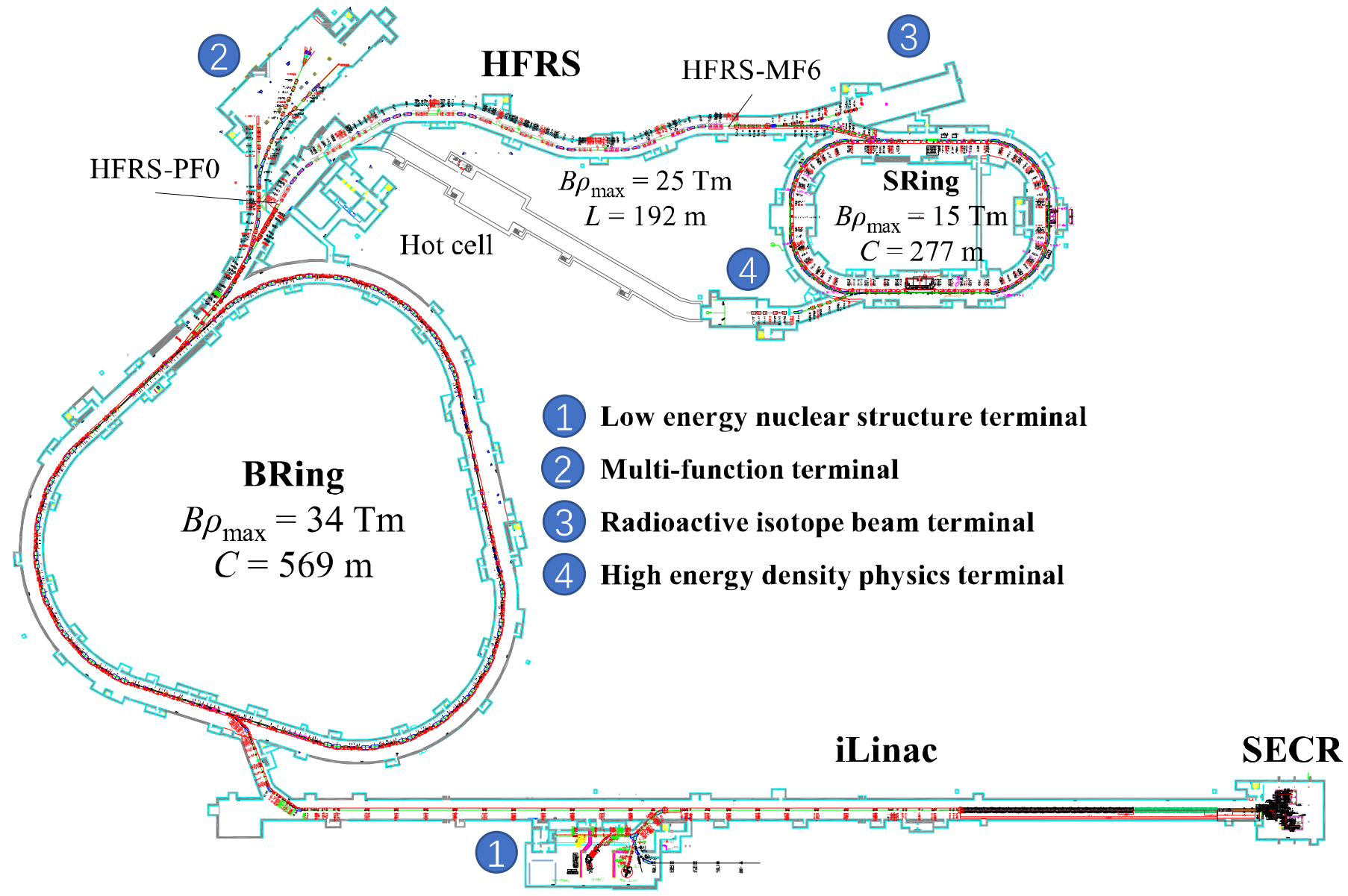

Figure 2. Layout of the HIAF.

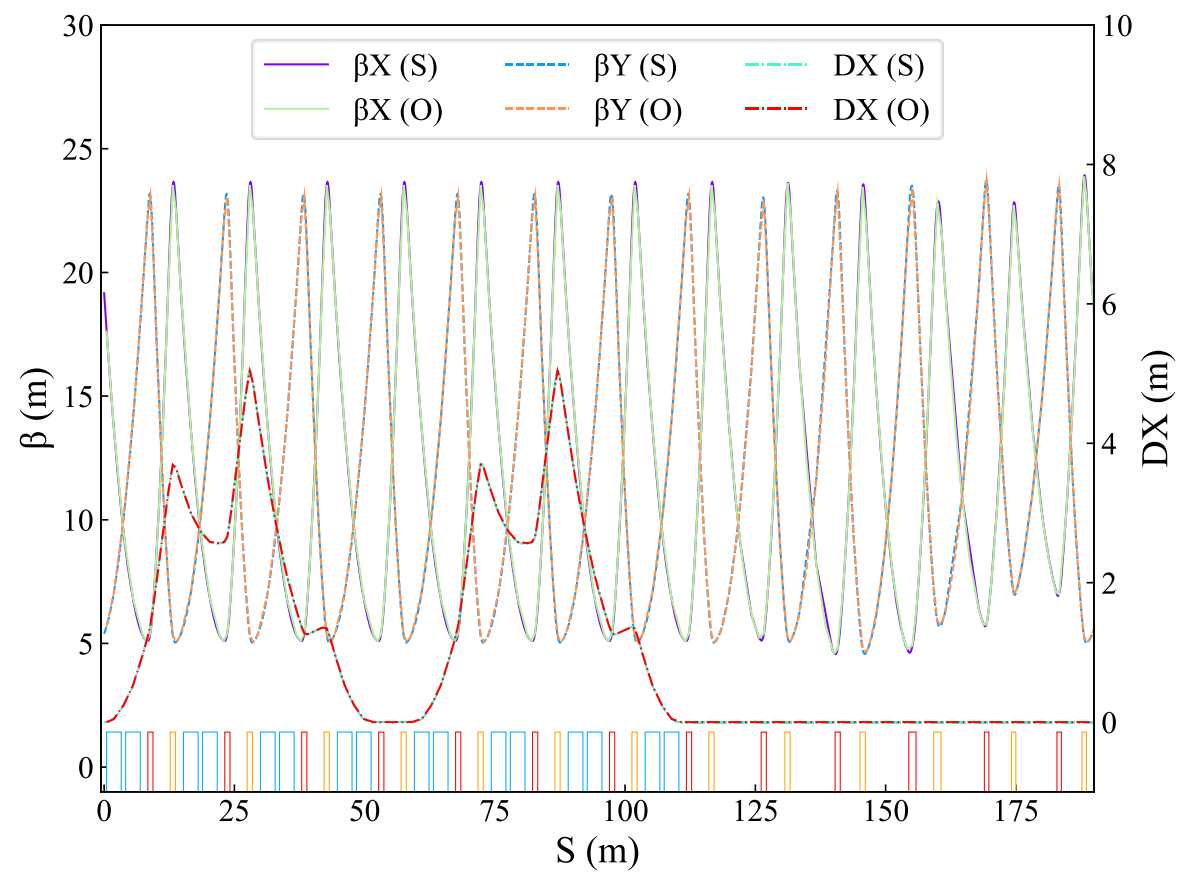

Figure 3. Betatron and horizontal dispersion functions of one super-period of BRing. S: Sliced optics. O: Original optics. Blue boxes represent dipole magnets, orange boxes represent horizontal focusing quadrupole magnets, and red boxes represent vertical focusing quadrupole magnets.

This paper conducts further simulation research on the high-precision sliced optics of BRing, focusing on three aspects: (1) closed-orbit distortion and variations in optical parameters caused by errors; (2) closed-orbit correction; (3) dynamic aperture. The sliced optics adopted in this paper is of the AC type [5], where both dipole magnets and quadrupole magnets are sliced. Moreover, the integral of the relative strength of each segment of a single magnet along the reference orbit equals the designed length. This model is employed by LACCS (Large-scale Accelerator Complex Control System), the control system of HIAF, and thus is consistent with practical application scenario.

## 2. Closed-orbit distortion and variations in optical parameters induced by errors

Table 1. Alignment errors (RMS) of magnets of BRing.

| Type | ΔX, ΔY, ΔS [mm] | Δφ, Δθ, Δψ [mrad] |
|---|---|---|
| Dipole Magnet | 0.2, 0.2, 0.5 | 0.2, 0.2, 0.1 |
| Quadrupole Magnet | 0.2, 0.2, 0.5 | 0.2, 0.2, 0.2 |

This section employs MAD-X [13] to investigate the closed-orbit distortion (COD) and variations in optical parameters of BRing induced by alignment errors and dipole magnet field errors.

COD refers to a phenomenon in accelerators, particularly circular accelerators like synchrotron radiation sources and colliders, where the actual particle trajectory deviates from the ideal closed orbit due to factors such as magnetic field errors and alignment errors. Its impact on accelerator optics covers core aspects including beam transport, focusing control, and performance limits. In linear optics, the primary causes of COD include field errors and alignment errors of dipole magnets, as well as alignment errors of quadrupole magnets.

The alignment errors of the magnets are listed in Table 1; all follow a Gaussian distribution with a 3σ cutoff. ΔX, ΔY and ΔS denote the magnet's alignment errors in the horizontal, vertical, and longitudinal directions, respectively. Δφ, Δθ and Δψ represent the magnet's rotation errors around the horizontal, vertical, and longitudinal axes, respectively. The relative field error (RMS) of the dipole magnets is 3.0E−4, which also adheres to a Gaussian distribution with a 3σ cutoff. The working points are set to $Q_x$ = 9.470 and $Q_y$ = 9.430. For both sliced optics and original optics, 1000 random error calculations are performed, with random numbers generated in the range [0, 99999999]. Statistics are collected for the maximum value and RMS of COD at various positions along the ring, as well as the maximum value and RMS of Twiss parameter variations. Additionally, statistics on variations in working points and transition gamma are conducted.

First, the COD is analyzed. As shown in Figure 4, the sliced optics exhibits slightly smaller horizontal COD with a marginally lower RMS value. The same trend applies to the vertical direction. It is evident that the sliced optics outperforms the original optics in uncorrected closed-orbit distortion.

In terms of the betatron function, as presented in Figure 5, the variation of the sliced optics is very close to that of the original optics. Regarding dispersion, as shown in Figure 6, the variation of the sliced optics is slightly smaller than that of the original optics, and the smaller dispersion variation in the bending section is conducive to beam transport. During the first beam commissioning on October 27, 2025, BRing successfully accelerated the $^{18}O^{6+}$ beam from 30 MeV/u to 2600 MeV/u. No closed-orbit correction was performed during this commissioning, and the beam intensity of the sliced optics reached the order of 1E+09, which was approximately 20% higher than that of the original optics.

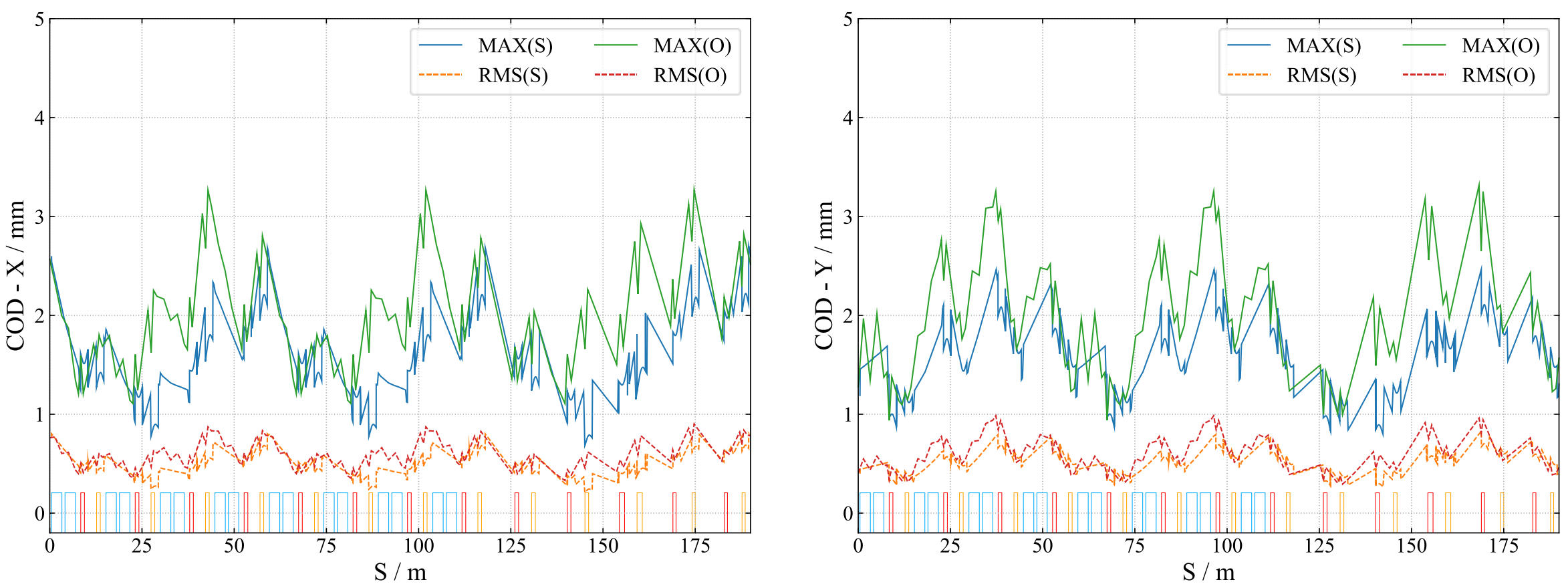

Figure 4. Closed-orbit distortion of one super-period.

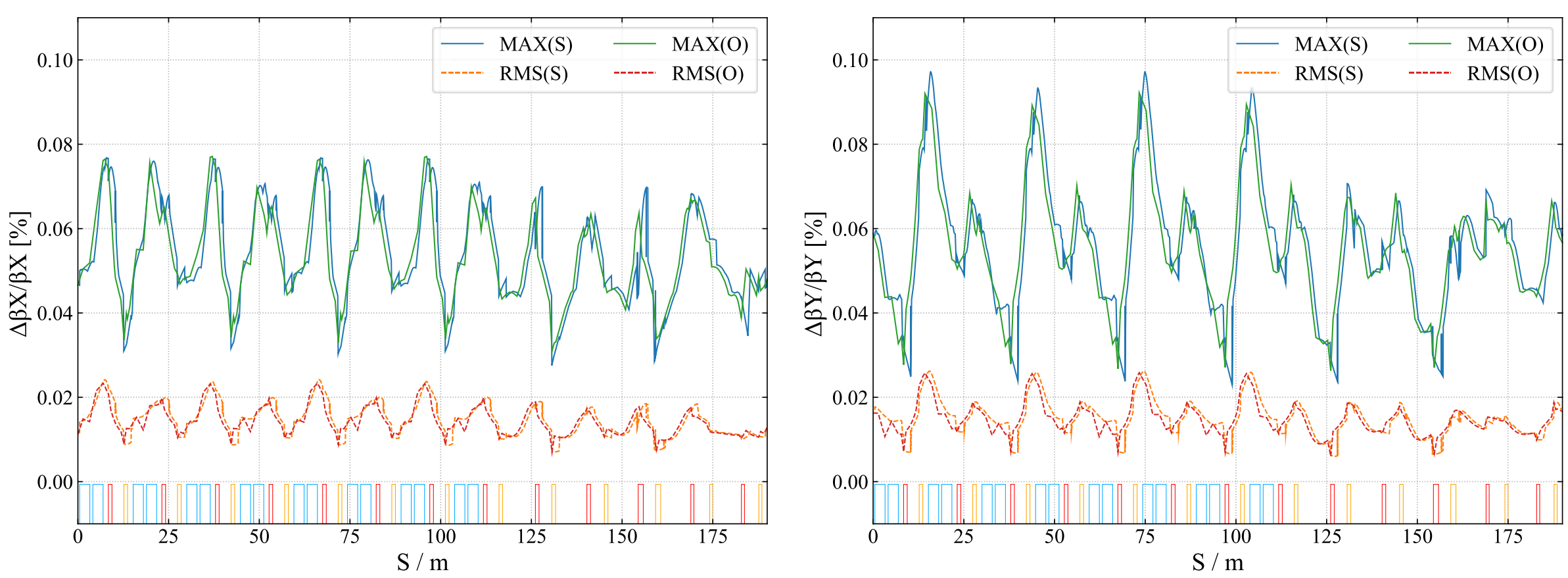

Figure 5. Variation in betatron function of one super-period.

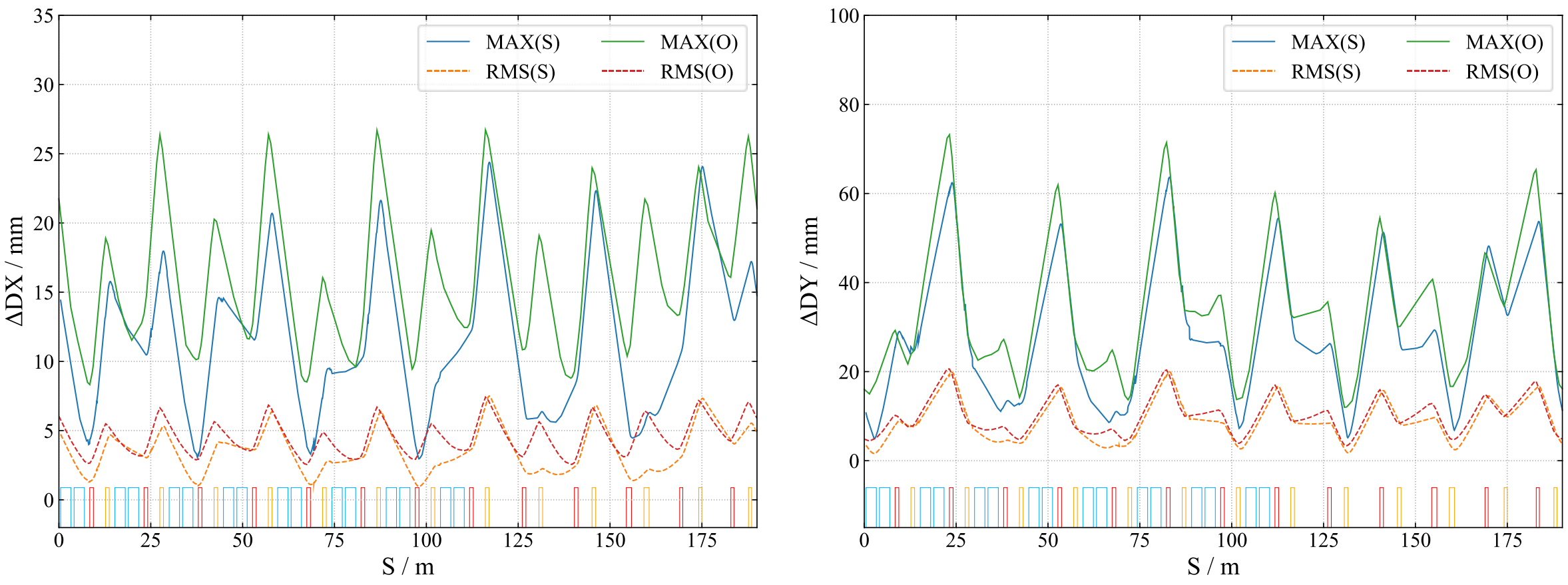

Figure 6. Variation in dispersion of one super-period.

Statistics on the variations of the working points and transition energy ($\gamma_t$) are presented in Figure 7 and Table 2. As indicated by the standard deviation, the variation in the transition gamma of the sliced optics is slightly smaller than that of the original optics. Regarding the working points, the variations of the two optics are very close.

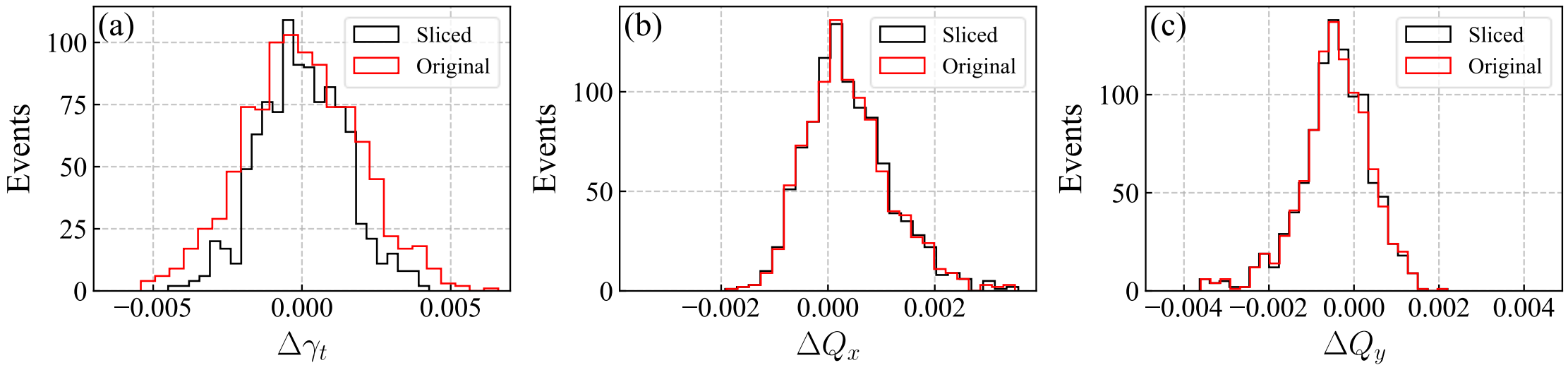

Figure 7. Variations of transition gamma and tunes.

Table 2. Standard Deviations of transition gamma and tunes.

| | $\Delta\gamma_t$ | $\Delta Q_x$ | $\Delta Q_y$ |
|---|---|---|---|
| Sliced Optics | 1.44E−03 | 7.97E−04 | 8.43E−04 |
| Original Optics | 1.92E−03 | 7.93E−04 | 8.38E−04 |

### 3. Closed-orbit correction

This section presents a simulation study on the closed-orbit correction of BRing. The simulation is performed using MAD-X, with calculation conditions consistent with those in Section 2, and all 79 correction magnets and 40 BPMs are employed. Of the correction magnets, 39 are used for horizontal orbit correction, 38 for vertical orbit correction, and the remaining 2 are bidirectional correction magnets. The layout of the correction elements is shown in Figure 1.

Figure 8 shows the closed-orbit distortion after correction. Following correction, the maximum COD of the sliced optics exceeds 1 mm only at a few individual positions, with the overall maximum value not exceeding 1.2 mm. In contrast, the COD of the original optics is slightly larger than that of the sliced optics at most positions, and its maximum value remains within 1.4 mm.

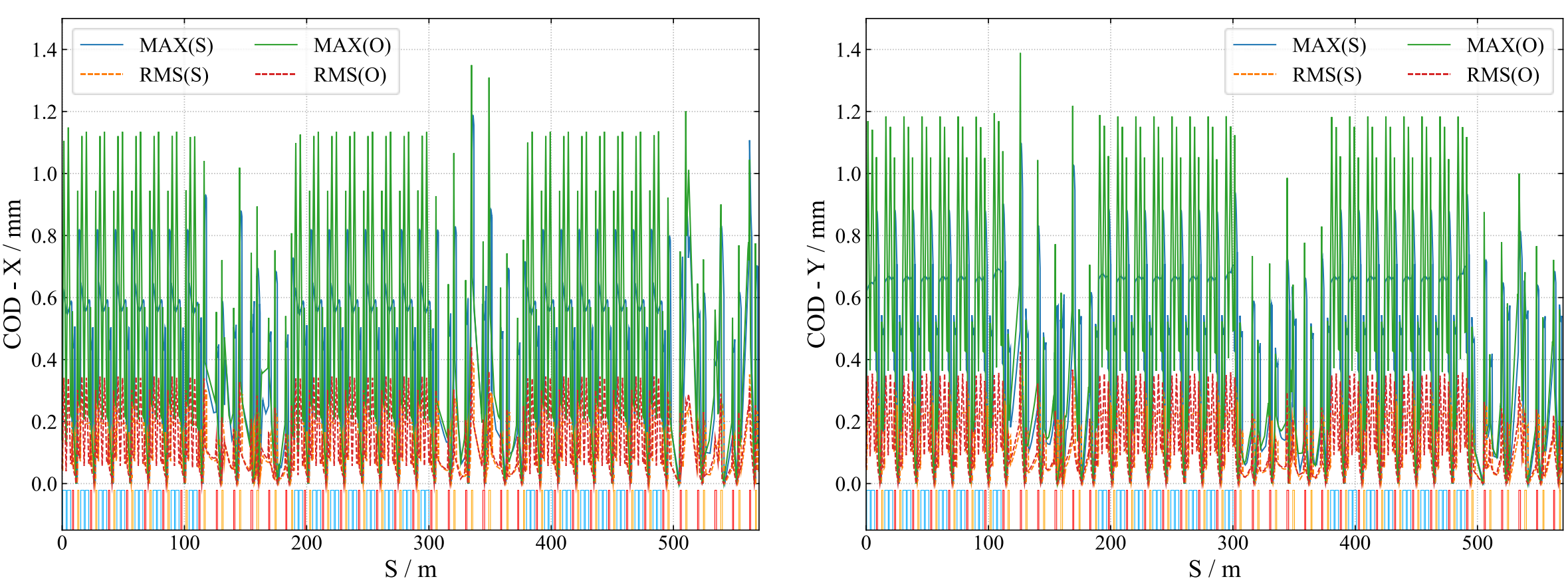

Figure 8. Closed-orbit distortion after correction.

In terms of the betatron function, after closed-orbit correction, the variation of the sliced optics is almost identical to that of the original optics, as shown in Figure 9, with little change compared to the uncorrected state. Similarly, the variation of the dispersion of the sliced optics is nearly identical to that of the original optics (as shown in Figure 10), but has been significantly reduced compared to the uncorrected state.

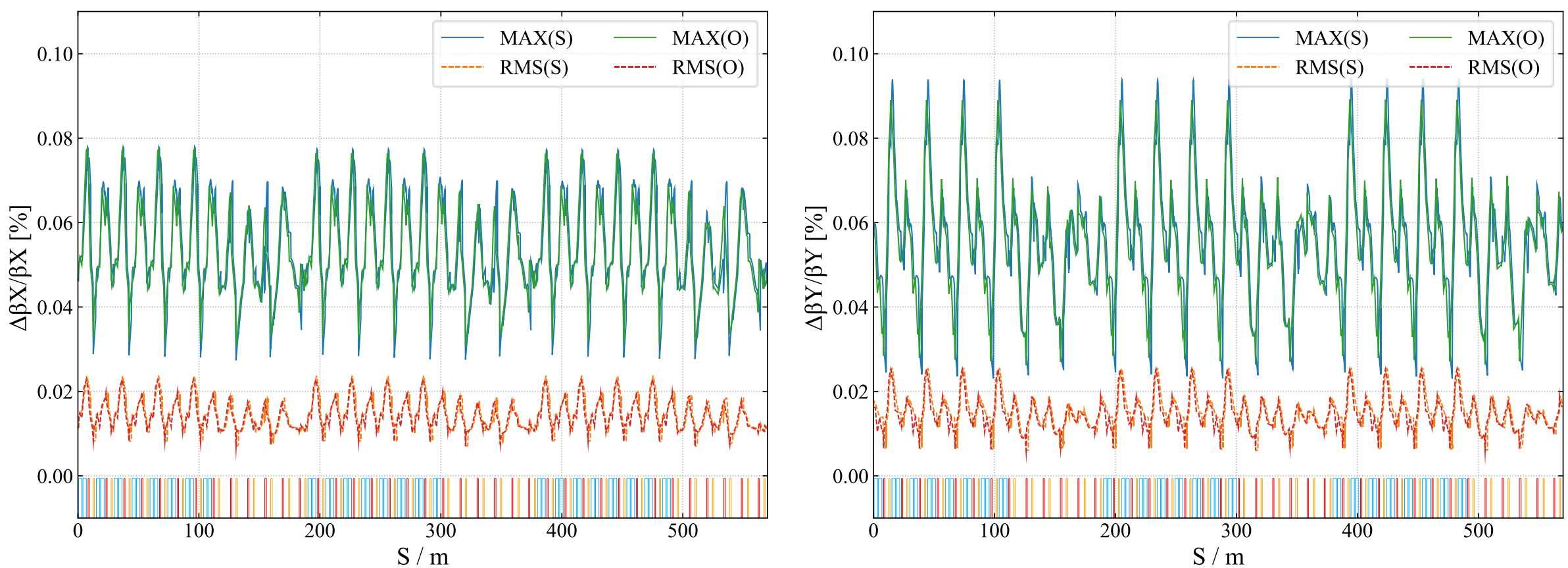

Figure 9. Variation in Betatron after correction.

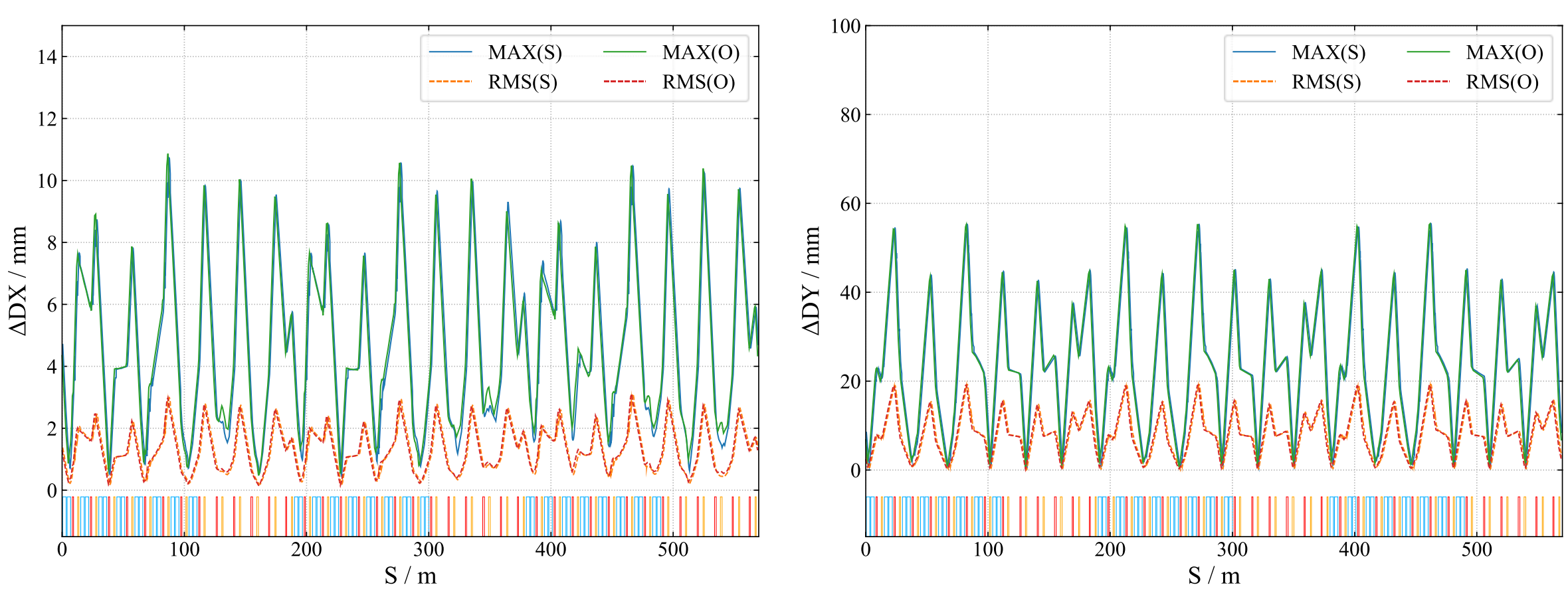

Figure 10. Variation in dispersion after correction.

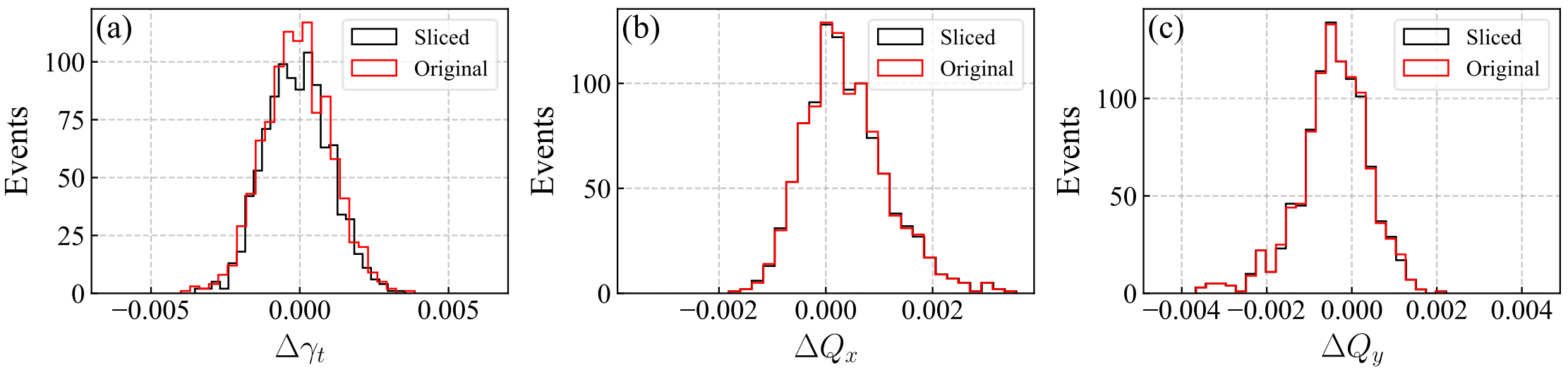

Figure 11. Variation of transition gamma and tunes after correction.

After correction, the variation of the transition gamma for the sliced optics is essentially consistent with that of the original optics. The same applies to the variation of the working points, as shown in Figure 11 and Table 3.

Table 3. Standard Deviations of transition gamma and tunes after closed-orbit correction.

| | $\Delta\gamma_t$ | $\Delta Q_x$ | $\Delta Q_y$ |
|---|---|---|---|
| Sliced Optics | 1.08E−03 | 7.91E−04 | 8.37E−04 |
| Original Optics | 1.11E−03 | 7.87E−04 | 8.32E−04 |

After closed-orbit correction, BRing's sliced optics exhibits smaller COD than the original, with a maximum value not exceeding 1.2 mm. The two optical models show high consistency across multiple core optical parameters: the sliced optics' betatron function variation is almost identical to the original's, with little deviation from its uncorrected state, and the variation patterns of transition gamma and working points are also essentially consistent. The only differential characteristic lies in the dispersion function: while the dispersion variation range aligns between the two schemes, the sliced optics achieves a significant reduction in dispersion compared to its uncorrected state.

## 4. Dynamic aperture

The dynamic aperture [14] can characterize the spatial range within which particles can oscillate stably in the transverse phase space in a circular accelerator, and it is one of the core performance indicators of a synchrotron. This section employs the Tao program [15] to study the dynamic aperture of BRing. The calculation conditions are as follows: (1) the machine's working points are set to $Q_x = 9.470$ and $Q_y = 9.430$; (2) magnet alignment errors and field errors are not considered; (3) the number of particle tracking turns is 1000.

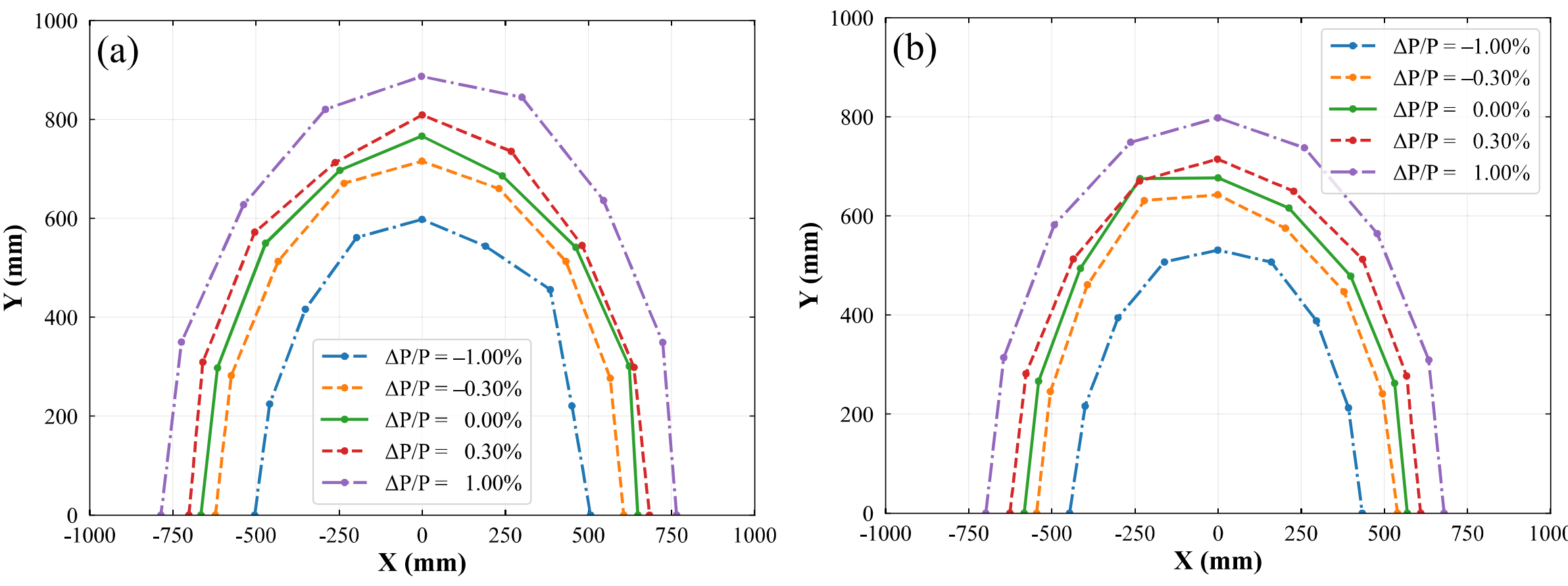

Figure 12. Dynamic apertures at injection without chromaticity correction. (a): Sliced optics. (b): Original optics.

First, the dynamic apertures of the sliced optics and original optics at the beam injection point and extraction point are calculated under five momentum spreads (0.00%, ±0.30%, and ±1.00%) without chromaticity correction. The calculation results are shown in Figures 12 and 13. It can be seen that at the injection point, the dynamic aperture of the sliced optics is significantly larger than that of the original optics; at the extraction point, the dynamic aperture of the sliced optics is also significantly larger than that of the original optics.

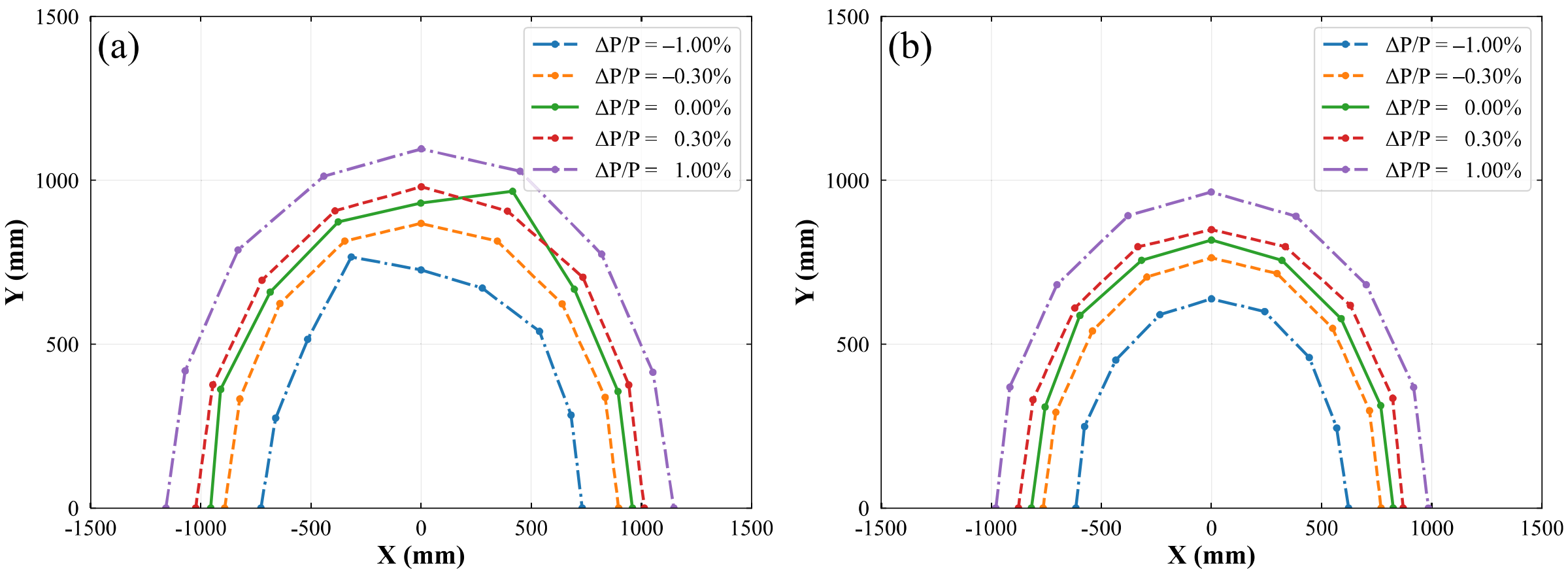

Figure 13. Dynamic apertures at extraction without chromaticity correction. (a): Sliced optics. (b): Original optics.

The chromaticity correction sextupole magnets of BRing are divided into two groups: one group is used to correct the horizontal chromaticity, and the other to correct the vertical chromaticity. The normalized gradients of the sextupole magnets within each group are consistent. Forty-eight sextupole magnets are used to correct both the horizontal and vertical chromaticity to 0. The normalized gradients of the sextupole magnets required for chromaticity correction in the two optics are comparable, as shown in Table 4.

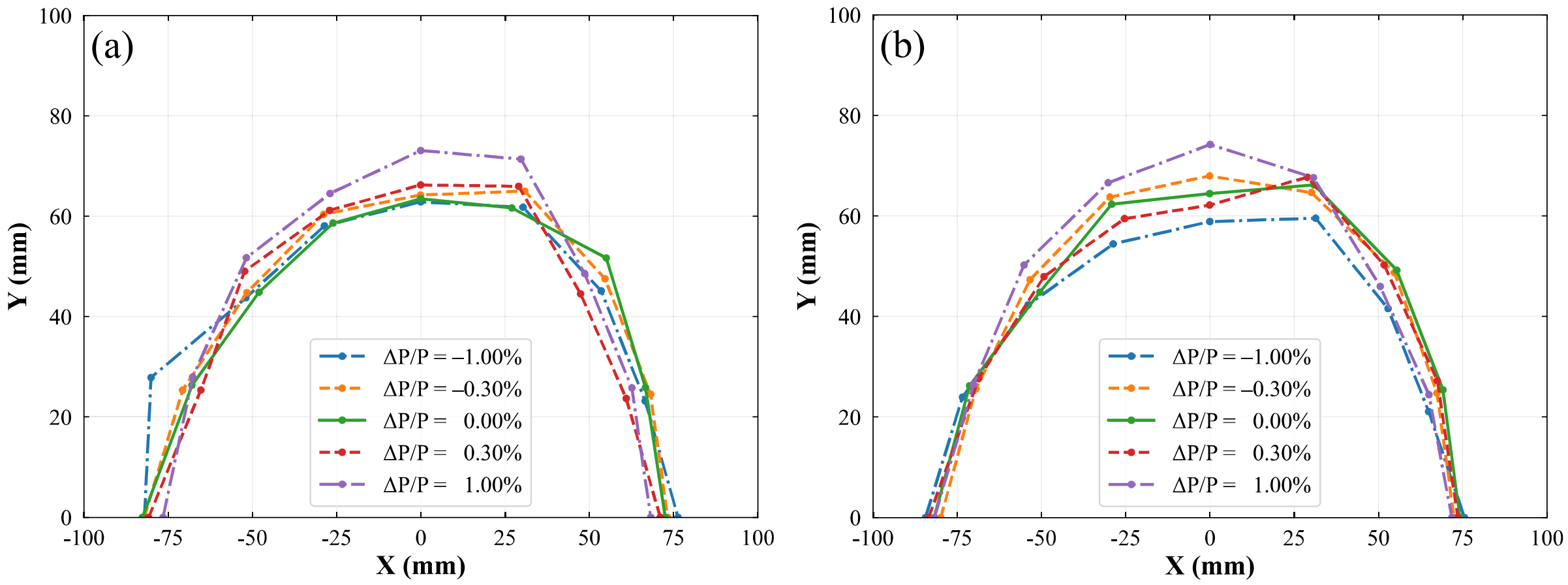

Figure 14. Dynamic apertures at injection with chromaticity correction. (a): Sliced optics. (b): Original optics.

The calculation results of the dynamic aperture of BRing with chromaticity correction are shown in Figures 14 and 15. It can be observed that the dynamic apertures of both optics are significantly reduced due to the strong sextupole magnetic fields introduced by chromaticity correction. After chromaticity correction, the dynamic aperture of the sliced optics is comparable to that of the original optics, whether at the injection or extraction point.

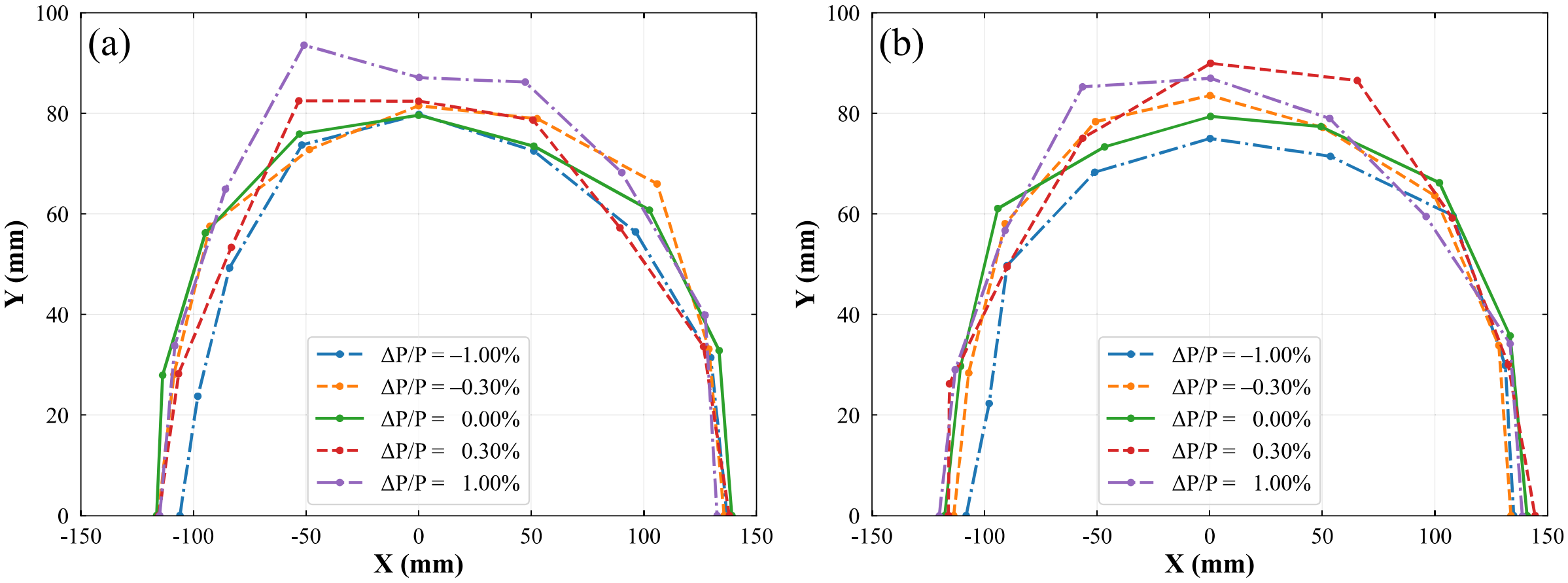

Figure 15. Dynamic apertures at extraction with chromaticity correction. (a): Sliced optics. (b): Original optics.

Table 4. Normalized gradients of the sextupole magnets for chromaticity correction.

| Type | Sliced Optics | Original Optics |
| --- | --- | --- |
| Horizontal Sextupole [$m^{-3}$] | 0.52505870635732199 | 0.52821722765464074 |
| Vertical Sextupole [$m^{-3}$] | –1.45757091097135949 | –1.46162551828544651 |

## 5. Summary

This paper conducts a further simulation study on the high-precision optics of HIAF-BRing.

First, it investigates the closed-orbit distortion (COD) caused by magnet alignment errors and dipole magnet field errors, as well as these errors' impacts on optical parameters. Without closed-orbit correction, the sliced optics outperforms the original optics in uncorrected COD. It shows slightly smaller horizontal and vertical COD with a lower RMS. This advantage is further validated by the first beam commissioning of BRing. The sliced optics also exhibits slightly smaller variations in dispersion and transition gamma compared to the original optics. Meanwhile, its betatron function variation and working points changes are very close to those of the original optics.

Then, a simulation study of closed-orbit correction for BRing is conducted. After closed-orbit correction, BRing's sliced optics exhibits smaller COD than the original, with a maximum value not exceeding 1.2 mm. The two optics demonstrate high consistency across core optical parameters (betatron function, transition gamma, working points), with the sliced optics showing little change from its uncorrected state. The only difference lies in dispersion: while its variation range aligns with the original optics, the sliced optics achieves a significant reduction in dispersion post-correction.

Finally, the dynamic apertures of the two optics of BRing are investigated. Without chromaticity correction, the dynamic aperture of the sliced optics is superior to that of the original optics both at the injection point and the extraction point. The sextupole magnetic fields required for chromaticity correction of the two optics are comparable. After chromaticity correction, the dynamic apertures of both optics are significantly reduced, and the dynamic aperture of the sliced optics is comparable to that of the original optics both at the injection and extraction points.

This study provides a more multi-faceted understanding of the high-precision optics of HIAF-BRing, which is conducive to maximizing the machine's performance and potential in future operations.

**Acknowledgements**

This work is supported by High Intensity heavy-ion Accelerator Facility (HIAF) project approved by the National Development and Reform Commission of China.